\documentclass[usenatbib]{basi}
\usepackage[british]{babel}
\usepackage[varg]{txfonts}
\def\degree{^\circ}
\usepackage[T1]{fontenc}

\begin{document}

\title[Rotating Radio Transients]{Rotating Radio Transients}

\author[E.~F.~Keane and M.~A. McLaughlin]
       {E.~F.~Keane$^1$\thanks{email:
           \texttt{ekeane@mpifr-bonn.mpg.de}} and
         M.~A.~McLaughlin$^{2}$\\ $^1$Max Planck Institut f\"{u}r
         Radioastronomie, Auf dem H\"{u}gel 69, 53121 Bonn,
         Germany. \\ $^2$Department of Physics, West Virginia
         University, Morgantown, WV 26506, USA.}

\pubyear{2011}
\volume{39}
\pagerange{\pageref{firstpage}--\pageref{lastpage}}

\date{Received 2011 August 22; accepted 2011 September 22}

\maketitle

\label{firstpage}

\begin{abstract}
Over the past several years, it has become apparent that some radio
pulsars demonstrate significant variability in their single pulse
amplitude distributions. The Rotating Radio Transients (RRATs),
pulsars discovered through their single, isolated pulses, are one of
the more extreme manifestations of this variability. Nearly 70 of these
objects have been found over the past several years in archival
and new pulsar surveys. In this review, we describe these searches
and their resulting discoveries. We then discuss radio timing
algorithms and the spin-down properties of the 19 RRATs with
phase-connected solutions. The spin-down parameters fall within the same range
as other pulsars, with a tendency towards longer periods and higher magnetic fields.
 Next we describe follow-up observations
at radio wavelengths. These show that there are
 periodic fluctuations in the pulse
detection rates of some RRATs and that RRATs in general have similar spectra to other pulsars.
X-ray detection has only been made for one RRAT, J1819$-$1458;
 observations have revealed absorption features and a bright X-ray nebula.
 Finally, we look to future
telescopes and the progress that
will be made with these in characterising and understanding the
Galactic RRAT population.
\end{abstract}


\section{Introduction}\label{sec:intro}
Pulsars do not often conform to the well-known textbook model: a
dependable `cosmic lighthouse' from which we can detect a train of
regularly-spaced pulses. Many are now known to exhibit very large
variations in radio emission, and in some cases even appear to switch
off for seconds, weeks, months or even years. To discover such
malfunctioning lighthouses we cannot use techniques dependent on there
being a reliable periodic signal. These sources are often much better
(or only) detectable via their occasional bright bursts of emission,
i.e. as `radio transients'. The eRRATic pulsars discovered in this way
are commonly dubbed Rotating Radio Transients, or `RRATs'.

In this review we first describe, in \S\ref{sec:psr_searches}, the
methods used in searches for pulsars and fast radio transients in
general. We then discuss, in \S\ref{sec:discovery}, the 2006 discovery
of the first 11 RRAT sources and the initial implications of
this. We proceed, in \S\ref{sec:new_discoveries}, with an overview of
the numerous subsequent searches for radio transients which have been
performed in the last 5 years. In \S\ref{sec:followups} we look at
the large body of follow-up observations which have been made,
primarily at radio wavelengths but spanning most wavelength ranges. We
conclude, in \S\ref{sec:discussion}, with a discussion of where these
erratic pulsars fit within the overall pulsar population.

\section{Pulsar searches}\label{sec:psr_searches}
Pulsar observations have focused, in recent times, on observing
frequencies of $\sim$1~GHz, where receivers with a few 100-MHz
bandwidth are used routinely. The radiometer equation then tells us
that such observing systems (on $\sim$100-metre class single dish
radio telescopes) can detect pulses of radio emission with peak flux
densities of $\sim$100$\sqrt{W_{\mathrm{ms}}}$~mJy, where
$W_{\mathrm{ms}}$ is the pulse width in milliseconds. The well-known
frequency-dependent dispersion a radio signal experiences as a result
of its journey through the interstellar medium is
$t_{\mathrm{delay}}=4.15({\rm DM}/\nu_{\mathrm{GHz}})\,\mathrm{ms}$,
where the dispersion measure ${\rm DM}=\int n_{\mathrm{e}} dl$ is the
integrated electron density along the line of sight in units of
pc~cm$^{-3}$ and $\nu_{\mathrm{GHz}}$ is the observing frequency in
GHz. Therefore, by channelising our bandwidth we can determine DMs
and, with a model for $n_{\mathrm{e}}(x,y,z)$ (e.g.\ \citealt{cl02}) in
the Galaxy, we can infer distances to sources. And so it is that we
can easily detect millisecond pulses with fluxes $\gtrsim$100~mJy and
estimate luminosities. For Galactic sources the
\textit{minimum} brightness temperature associated with such transient
radio signals is $4\times10^{23}\,\mathrm{K}
(S_{\mathrm{Jy}}D_{\mathrm{kpc}}^2)/(\nu_{\mathrm{GHz}}W_{\mathrm{ms}})$,
where $S_{\mathrm{Jy}}$ is the peak flux in Jy and $D_{\mathrm{kpc}}$
is the distance in kpc. Radio pulsars have the highest brightness
temperatures of all known radio transient sources, with most emitting
single pulses in the range
$10^{17}-10^{25}$~K, with the giant pulses from the Crab pulsar as
high as $10^{38}$~K \citep{hkwe03}. Additionally, millisecond pulse
widths, by causality, imply \textit{maximum} source sizes of
$300W_{\mathrm{ms}}$~km. Clearly the emission mechanism responsible
for such signals involves non-thermal, coherent emission from compact
objects.

\subsection{First discoveries}
Even though thought to exist since the 1930s~\citep{bz34,ov39} neutron
stars were seen as `theoretical', i.e. difficult/impossible to
detect. It was not until the serendipitous discovery of the first
pulsar in 1967~\citep{hbp+68} that neutron stars were promoted to the
forefront of astrophysical research. Although until this time radio
astronomers had not observed at sufficiently high time resolution for
their detection, this soon became commonplace. The first pulsars were
discovered by visual inspection of the time series from large radio
telescopes, at first with no correction for dispersion, but this was
quickly incorporated. The first $\sim$30 pulsars (see \S~3.4 of
\citealt{ls04} and references therein for a historical review) were
discovered in this way, amongst which the most noteworthy is the Crab
pulsar~\citep{sr68}. The short period of 33~ms (although not measured
in the discovery observations) solidified the neutron star
interpretation (over the white dwarf interpretation) of pulsars.

\subsection{Periodicity searches}
It was quickly realised that Fourier domain searches would enable the
detection of pulsars whose individual pulses were not bright enough to
be detectable. For a train of pulses of duty cycle $\delta=W/P$
(i.e. width $W$ recurring at a period of $P$) an improvement in
sensitivity by a factor of $\sqrt{(1-\delta)/\delta}$, or of
$\sim1/\sqrt{\delta}$ for narrow pulses, can be made. Fourier
searches, wide DM-range searches, and the use of extended bandwidths
resulted in the discovery of many more
pulsars~\citep{mld+96,mlc+01}. Faster time sampling allowed the
discovery of millisecond pulsars (MSPs)~\citep{bkh+82} and
sophisticated techniques for detecting binary systems were
developed~\citep{fcl+01,ran01}.

The standard pulsar search pipeline had become (see \S6 of
\citealt{lk05} for an extensive overview):

\begin{enumerate}

\item Observe at frequencies $\sim$1~GHz, a trade off between the
  effects of scattering, dispersion and sky temperature, which would
  favour higher frequencies, and typical steep pulsar spectral
  indices~\citep{mkk+00}, which would favour lower frequencies (see
  e.g.\ \citealt{mlc+01}).

\item Use as wide a bandwidth as possible to increase sensitivity,
  with as many frequency channels as possible to reduce intra-channel
  smearing to enable detection of MSPs (see
  e.g.\ \citealt{kjs+10,bbb+11}).

\item Choose a DM range to search (perhaps based on the sky position
  and the expected maximum DM in that direction) and for each DM in
  that range, dedisperse the data by shifting the frequency channels
  to compensate for frequency-dependent delays. The spacing of the DM
  trial values, or `channels', is selected so that the error due to an
  incorrect DM is no smaller than the intrinsic dispersion smearing
  over a single frequency channel.

\item Fourier transform these dedispersed time series and apply
  appropriate filters in the frequency domain to detect periodic
  signals. For each significant signal, fold the original dedispersed
  data at the candidate period (before tuning the DM and period
  values for an optimal detection). An acceleration search for binary
  pulsars adds either an extra layer to the nested search loop, over
  the range of acceleration values, or increases the complexity of the
  Fourier domain search at each DM trial.

\end{enumerate}

\subsection{Single pulse searches}
After the initial pulsar discoveries, and the switch to the `standard'
searching procedure, it was not standard to search the dedispersed
time series in the time domain for bright individual pulses (with a
few exceptions, e.g.\ \citealt{Nic99}). However, there are a number of
reasons why it is sensible to add such a `single pulse search' to the
pipeline. Firstly, we should exhaust all possible avenues to
maximise the yield of our search, especially as it is
computationally very inexpensive to add single pulse searches to the
standard search setup. Secondly, it is not only quick, but effective,
as there are many pulsars which are more significantly, or indeed
only, detectable via single pulse searches.

It is straightforward to show~\citep{mc03} that the ratio of
signal-to-noise ratios of FFT and SP searches (the so-called
`intermittency ratio') is given by:
\begin{equation}\label{eq:r}
  r=\frac{(S/N)_{\mathrm{SP}}}{(S/N)_{\mathrm{FFT}}}=\frac{A}{\sqrt{N}}\frac{S_{\mathrm{peak}}}{S_{\mathrm{ave}}}
  \; ,
\end{equation}
where $A$ is a product of constants and is $\approx 2$, $N$ is the
number of pulse periods during the observation, and
$S_{\mathrm{peak}}$ and $S_{\mathrm{ave}}$ are the peak and average
pulse flux densities, respectively. Depending on the pulse amplitude
distribution and the observation specifications, a pulsar may be more
effectively (or only) detectable in either a single pulse (SP) or a
Fourier domain (FFT\footnote{Although `FFT' denotes Fast Fourier
  Transform, it is used here, as is common, as a synonym for all
  searches which rely on the time-averaged emission.}) search. For
longer observations there is a higher probability of a pulsar emitting
a very bright single pulse (i.e. $S_{\mathrm{peak}}\gg
S_{\mathrm{ave}}$) and hence increasing the likelihood of detection in
a single pulse search. However in the limit of high $N$ the
periodicity search will become more effective. Many pulsar amplitude
distributions have a `sweet spot', i.e. a peak in $r$ as a function of
$N$, between where these competing effects dominate (see
\citealt{kea10a} for an examination of several families of
distributions).

\begin{figure}
  \begin{center}
    \includegraphics[scale=0.4,angle=-90]{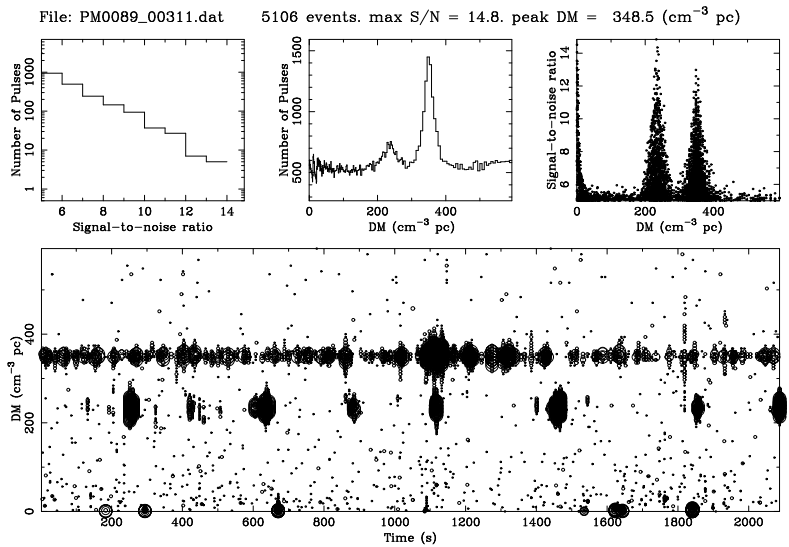} 
  \end{center}
  \caption{\small{A typical diagnostic plot from a single pulse
      search, in this case produced using the \textsc{sigproc}
      software suite. The top line gives some summary information and
      the top three panels show, from left to right, the number of
      pulses as a function of signal-to-noise ratio, the number of
      pulses as a function of DM, and the signal-to-noise as a
      function of DM. The bottom panel shows all detected events
      (above some user-defined threshold, in this case $5$ standard
      deviations above the mean) in the DM-time plane plotted as
      circles whose size is proportional to signal-to-noise.
 This particular observation is from the Parkes
      Multi-beam Pulsar Survey. It is 35-minutes in length and has been
      searched for significant pulses in over 300 trial DM values
      covering a range of $0-2100\,\mathrm{cm^{-3}}\,\mathrm{pc}$; DM values up to 600\,$\mathrm{cm^{-3}}\,\mathrm{pc}$
are shown here. Two
      pulsars (J1840$-$0809 and J1840$-$0815) can be clearly seen as numerous significant
      events at constant DM values.
The properties of PSRs~J1840$-$0809 and J1840$-$0815 are very similar, with periods, DMs, and 1400-MHz fluxes of
0.95~s, 350~pc~cm$^{-3}$, and 2.3~mJy and  1.09~s, 233~pc~cm$^{-3}$, and 1.4~mJy,
respectively.
However, the degrees of modulation of these pulsars are incredibly different. One can see by
eye that  PSR~J1840$-$0815 has a higher
      `modulation index', i.e. $\Sigma_{\mathrm{i}}^{\mathrm{N}}
      (S_{\mathrm{i}}-S_{\mathrm{ave}})^2/S_{\mathrm{ave}}^2$, and if
      it were more distant (equivalent to reducing our sensitivity) we
      would see just the very brightest events and the source may
      only be detectable in a single pulse search. The pulses at a DM of zero are due to RFI.}}
  \label{fig:t-dm}
\end{figure}

Single-pulse searches are quite straightforward. They involve
searching each dedispersed time series for significant peaks, an
exercise in matched filtering. Typically the dedispersed time series
is searched for peaks, i.e. pulses of width $t_{\mathrm{samp}}$, the
sampling time, then `decimated' in time (i.e. adjacent time samples
are added) by a factor $n$ and searched for peaks, i.e. pulses of
width $n\times t_{\mathrm{samp}}$. Usually decimation is done in
factors of  two (with more accurate  fits made later to
obtain more precise pulse widths). For example, the Parkes Multi-beam
Pulsar Survey (PMPS), with $t_{\mathrm{samp}}=250\;\muup$s, was recently
searched for pulses with pulse-widths ranging from $2^0$ to $2^9$
times this sampling time~\citep{kle+10}. This is equivalent to
convolving the time series with box-cars of width $n\times
t_{\mathrm{samp}}$ with $n=1,2,4,8,\ldots,512$. Once significant
events over a range of widths have been recorded for a range of DM
values, a plot like that shown in Fig.~\ref{fig:t-dm} is examined
for significant events at a constant non-zero DM.

Further techniques are also often employed to remove signals that are
due to terrestrial interference, which can obscure astrophysical
signals of interest, such as `zero-DM subtraction'~\citep{ekl09} or
coincidence matching when using multi-beam receivers
\citep{dcm+09,kle+10,bbj+11}. The software packages usually used (see
\citet{mcl11} for a discussion) to search pulsar survey data are
\textsc{sigproc}\footnote{\texttt{http://sourceforge.net/projects/sigproc}},
\textsc{swinlegion}\footnote{\texttt{https://github.com/swinlegion}}
(an offshoot of \textsc{sigproc} but with many extra features) and
\textsc{presto}\footnote{\texttt{https://github.com/scottransom/presto}}. All
of these packages are freely available.

\section{RRATs: discovery}\label{sec:discovery}
With the aim of discovering pulsars more easily detectable via their
single pulses, or indeed any transient radio events, \citet{mll+06}
searched the entire PMPS \citep{mlc+01} with the search algorithm
described in \citet{cm03} and as outlined above. To date, the PMPS is
the most successful pulsar survey ever performed, with a yield of over
800
pulsars~\citep{mlc+01,mhl+02,kbm+03,hfs+04,fsk+04,mll+06,lfl+06,kel+09,ekl09,eat09,kle+10,emk+10,kkl+11}. A
decade later this number is still growing as the survey is currently
being reprocessed with a number of new algorithms.

\subsection{Eleven new sources}
\citet{mll+06} identified 11 new sources which were not detectable in
PMPS periodicity searches, resounding proof of the value of single
pulse searches. They were characterised by repeated bursts at a
constant DM and were at first referred to as Repeating Radio
Transients. However, before the initial publication, with follow-up
observations of between 8 and 30 hours, and the detection of between 4
and 229 pulses from each source, it became clear that there were
underlying periodicities which ranged from 1--7 seconds, typical
neutron star rotation periods. Thence these sources were recognised to
be rotating neutron stars and re-named Rotating Radio Transients
(RRATs). Further evidence of the neutron star nature of these 11
sources included the facts that the pulse widths of a few milliseconds
implied (by causality) very compact sources (and were close to the
neutron star dynamical timescale), their inferred minimum brightness
temperatures were similar to those of the known pulsars, positive
period derivatives with typical pulsar-like values were measured for
three sources. Additionally, shortly after the discovery, one of the
sources was seen in thermal X-rays as expected for a cooling neutron
star \citep{rbg+06}.

\subsection{First ideas, questions, and implications}
The discovery of the first 11 RRATs in the PMPS was followed by a
number of papers presenting theories to explain the irregularity of
the detected radio emission. Radio pulses were detected at rates as
low as once per three hours  to as frequently as one every few
minutes. The question of interest to many was whether the sources were
truly `off' (or to use pulsar parlance, `nulling') during the rotation
periods where no pulses were detected or whether they had some weak
underlying emission. The non-detection in periodicity searches of
PSR~J1819$-$1458, the source with the brightest and the highest number
of detected pulses, implied a lower limit on
$S_{\mathrm{peak}}/S_{\mathrm{ave}}$ of $\sim$200~\citep{mll+06}. For
J1317$-$5759 this limit was a factor of $\sim$125. Such numbers are
not unheard of, e.g.\ in PSR~B0656$+$14 this ratio is $\sim$420,
leading \citet{wsrw06} to point out that any pulsar with similar
pulse-to-pulse modulation would be detected as a RRAT if at a
sufficient distance from Earth. If B0656$+$14 were moved $\sim$12
times further away to a distance of $3.5$~kpc it would only be
detectable in a single pulse search (and we note that the distances of
the 11 PMPS RRATs ranged from $\sim$2--7~kpc). For the other nine
sources there simply were not enough data to determine limits on
$S_{\mathrm{peak}}/S_{\mathrm{ave}}$. Nonetheless, many authors
proposed different mechanisms for how switching `on' and `off'
pulsar-like emission could occur, applying these to the 11 PMPS RRATs
assuming they were `off' when not detected. These ideas involved, to
name but a few, re-activating ceased pulsar emission due to transient
disturbances in supernova fall-back discs~\citep{li06}, the release of
plasma trapped in radiation belts~\citep{lm07} or the interaction with
proposed asteroidal material around the star~\citep{cs08}.

Another implication from the discovery of the 11 PMPS RRATs was that
there remained a large number of undiscovered neutron stars of this
type, i.e. those with sporadic and/or highly-modulated radio emission
which are more easily or only detectable in single-pulse
searches. Considering the likelihood of bursts being detected in the
35-minute pointings of the PMPS, the number and distribution (e.g.\ in
`radio luminosity' $SD^2$), the extent of contamination of the survey
due to interference and beaming effects, \citet{mll+06} estimated the
total number of such sources in the Galaxy to be larger than the
`normal' pulsars (i.e. those much more easily detectable in
periodicity searches) by as much as a factor of four. This of course
was an estimate based on a small number of sources and on a number of
parameters whose values were not well-determined. However, even if
over-estimated by an order of magnitude this expected number is still
a very large fraction of the overall pulsar population. Another
suggestion made in the initial discovery paper was that these sources
represented a distinct and previously unknown population of bursting
neutron stars.

An estimate of the numbers and birthrates of Galactic neutron stars
was made by \citet{kk08} to examine the implications of this. The
conclusions were that if the neutron star classes, not just the
`normal' pulsars and the new PMPS RRATs, but also the XINS (X-ray
Isolated Neutron Stars, see e.g.\ \citealt{k08})\footnote{These sources
  have also been referred to as XDINS (X-ray Dim Isolated Neutron
  Stars), INS (Isolated Neutron Stars), the `Magnificent Seven', and RQNS
  (radio quiet neutron stars) in various publications.}, magnetars~\citep{td96},
and CCOs or Central Compact Objects~\citep{hg10} were distinct, then the observed Galactic
supernova rate (which is presumably greater than or equal to the
neutron star birthrate) seemed to be insufficient. Identifying the
populations with each other could remove this problem, leading to the
suggestion that the PMPS RRATs and `normal' pulsars were part of the
same underlying population.

The discovery of the first 11 PMPS RRATs raised questions regarding
the pulsar emission mechanism(s) as well as our knowledge of how many
neutron stars there are in our Galaxy. To address these questions a
large observational campaign of the 11 known sources was necessary, as
well as systematic searches to identify the large predicted number of
sources. Below, in \S\ref{sec:new_discoveries}, we describe the
numerous successful searches for new RRATs in various radio surveys,
and then, in \S\ref{sec:followups}, we describe the follow-up
observation campaigns both at radio and other wavelengths.

\section{The search continues}\label{sec:new_discoveries}

There are several ways in which one might define a RRAT. One could argue that they are
simply one extreme of the  normal pulsar intermittency spectrum and therefore require no definition. However,
for the purposes of this review and other papers, one could simply call RRATs those pulsars discovered
only through a single-pulse search, as in \citet{mll+06}. Or, the distinguishing characteristic of a RRAT  could
be a pulsar from which we see predominantly isolated pulses (see e.g.\ \citealt{bb10}).
These are both viable but for the purposes of this review, we propose:

\textit{A RRAT is a repeating radio source, with underlying
  periodicity, which is more significantly detectable via its single
  pulses than in periodicity searches.}

A pulsar is then labeled a `RRAT' if, for the observing
specifications concerned, $r>1$ (see Equation~\ref{eq:r}). A pulsar
does not have to exhibit long timescale `nulling' (i.e. sustained periods where radio
emission is `off') to satisfy this criterion. Clearly the definition
is rather arbitrary as a pulsar may be a
`RRAT' in one survey but not in another --- hence the use of `PMPS
RRAT' etc. At the time of writing, according to this definition, 61 pulsars have been identified
as RRATs \citep{mll+06, hrk+08, skdl09, kle+10, bb10, rub10, kkl+11, bbj+11}\footnote{Different numbers of RRATs are reported in several
  publications, for a number of reasons. Firstly, sources do not have
  a constant value of $r$ between observations with different
  specifics, or even between identical observing setups as the
  sampling of the pulsars' amplitude distribution functions will
  always yield different $S_{\mathrm{peak}}/S_{\mathrm{ave}}$
  values. This is a natural consequence of the loose `definition' of
  RRATs and we re-emphasise that the label should not be taken too
  strictly. Secondly, there have been different, even looser,
  `definitions' used by different authors (the one given here
  follows \citealt{kkl+11}), none of which have been strictly applied,
  e.g.\ if sources must be completely undetectable in FFT searches then there will be fewer RRATs
   and if it is simply the discovery method which
  decides the label then there are \textit{many} more pulsars which
  should be labeled so, e.g.\ the Crab pulsar. } in
various surveys, which we now review in turn.
 This count includes all repeating sources discovered in the above papers through single-pulse searches.

\subsection{Archival Parkes surveys}
The PMPS, wherein the original 11 RRAT sources were discovered,
surveyed the Galactic plane in the longitude range $260\degree$ to
$50\degree$ and latitude range $|b|<5\degree$ at a central observing
frequency of $1374$~MHz spanned by $96\times3$~MHz channels sampled
every $250\,\muup$s. Two further surveys were performed with the Parkes
telescope at intermediate and high Galactic latitudes of
$5\degree<|b|<30\degree$~\citep{ebsb01,jbo+09} in the same longitude
range as the PMPS. The surveys used almost identical survey
specifications to the PMPS, except for a faster sampling time of
$125\,\muup$s and shorter $4.4$-minute pointings. \citet{bb10} analysed
both of these surveys, performing single pulse searches and
discovering 14 new sources.

With an estimated $\sim$50\% of the expected RRAT sources in the PMPS
obscured by radio frequency interference~\citep{mll+06} and the
development of some algorithms for its effective mitigation (see
e.g.\ \citealt{ekl09}), \citet{kle+10,kkl+11} re-analysed the PMPS,
discovering 17 new sources. Another notable search of the Parkes
archives was performed by \citet{lbm+07} and resulted in the discovery
of the so-called `Lorimer Burst', a 5-ms duration burst from an
unknown apparently extra-galactic source. With many such signals
expected to be found in pulsar surveys, the DM range of single pulse
searches now commonly reaches well beyond the `edge of the Galaxy',
i.e. the highest DM expected for a Galactic source based on some model
for the electron density distribution (e.g.\ the NE2001 model of
\citealt{cl02}). Indeed one of the bursts reported in \citet{kkl+11} is
also possibly extra-galactic in origin.

All of these high-impact scientific discoveries show the value of
searching archival datasets (even those which have been examined many
times already). At present there remain many archival Parkes surveys
which have not been subjected to single pulse searches. These data,
which are publicly available, are a source with huge potential for
discovery. The recent and ongoing efforts by \citet{hmm+11} to make
the Parkes pulsar archives easily accessible to all online will no
doubt help in the endeavour to exhaustively search all available data
for maximum scientific output.

\subsection{The HTRU}
The latest and ongoing pulsar search occurring at the Parkes Telescope
is the High Time Resolution Universe (HTRU, aka HTRU South)
survey~\citep{kjs+10}. This survey is being performed at a central
frequency of 1352 MHz with a usable (total) bandwidth of 340 (400)
MHz. The frequency and time resolution are $390$~kHz and $64\,\muup$s
respectively, $\sim$8 and $\sim$4 times higher than that of the
PMPS. This has allowed the discovery of intrinsically narrower pulses,
enabling more distant MSPs to be identified (see
e.g.\ \citealt{bbb+11}). The greater sensitivity, coupled with the fact
that the HTRU survey covers areas outside of the PMPS sky, means the
expected yield of new pulsars is high, numbering around
400~\citep{kjs+10}. Recently \citet{bbj+11} have performed a single
pulse search on the HTRU data. The search covered $0\%$, $23.5\%$ and
$0.4\%$ respectively of the HTRU low-, intermediate- and high-latitude
pointings and resulted in the discovery of 11 new sources. The twin
northern hemisphere HTRU survey will be performed with the Effelsberg
Telescope in Germany and has recently commenced.

\subsection{Low-frequency Green Bank Telescope searches}
Single-pulses searches have been incorporated in several low-frequency
pulsar surveys with the Green Bank Telescope (GBT).  The first
low-frequency survey \citep{hrk+08} used a central frequency of
$350$~MHz with $50$-MHz bandwidth, 1024 frequency channels, and
$81.92$-$\muup$s sampling. This low frequency, combined with the
sensitivity of the GBT, makes this survey very sensitive to faint,
nearby pulsars. The survey covered roughly 1000 square degrees at $|b|
< 5.5\degree$ and $75\degree < l < 165\degree$ and resulted in the
discovery of 33 new pulsars, five of which were found through the
single-pulse search. Two of these pulsars are very sporadic. One is
undetectable for nearly 97\% of the time and another, which \citet{hrk+08} classify as the only RRAT detected in their survey,
shows individual
bright bursts separated by 100s -- 1000s of seconds.

A second, similar GBT survey was undertaken in the (Northern) summer
and autumn of 2007. During this time, $10300$ deg$^2$ between
declinations of $-21\degree$ and $+26\degree$ were covered in
`drift-scan' mode during the GBT track repair, with an effective
integration time of 150~s.  A central frequency of $350$~MHz, $50$-MHz
bandwidth, 2048 frequency channels, and $81.92$-$\muup$s sampling time
were used. A total of 26 pulsars were discovered, including one object
through the single-pulse search. This pulsar is, however, typically
detectable through an FFT. A portion of the data (20 Terabytes, or
1500 deg$^2$) has been set aside for analysis by high-school students
in the Pulsar Search Collaboratory (PSC) project \citep{rhm+10}.

Encouraged by these successes, a new $350$-MHz GBT survey of the
Northern celestial cap is currently being undertaken, with the primary
goal of searching for pulsars to include in pulsar timing arrays for
gravitational wave detection. This survey will cover declinations
greater than $38\degree$ at $350$~MHz with two-minute integrations
using $100$-MHz bandwidth, 4096 frequency channels, and 81.92-$\muup$s
sampling time. Given the number of sources detected in single-pulse
searches in previous GBT surveys, this new survey is expected to
discover dozens of RRATs.

\subsection{The PALFA survey}
The PALFA (Pulsar Arecibo L-band Feed Array) survey uses a seven-beam
receiver on the 305-m Arecibo telescope to cover the region $|b| <
5\degree$ and $40\degree < l < 75\degree$, $170\degree < l <
210\degree$ with $300$-MHz of bandwidth, 256 frequency channels,
64-$\muup$s sampling time, and 265-s integrations \citep{cfl+06}. This
survey has thus far resulted in 56 pulsar discoveries, with seven of
those made through a single-pulse search~\citep{dcm+09}. Several of
these objects are very intermittent, with less than 10 pulses detected
in many hours of follow-up observations. This survey should ultimately
discover several hundreds of new pulsars, with tens of these
identified as RRATs.

\subsection{Other}
In addition to the above sources, the discovery of one RRAT source at
the Puschino Observatory in Russia was reported by
\citet{skdl09}. Furthermore the Westerbork Synthesis Radio Telescope's
8gr8 survey~\citep{jsb+09} of the Cygnus region at $328$~MHz has
recently reported the detection of four RRAT sources~\citep{rub10}.

\section{Follow-up work}\label{sec:followups}
Many of the sources found in the surveys described above in
\S\ref{sec:new_discoveries} have been subject to extensive follow-up
observations, which we now describe.

\subsection{Radio}
When new pulsars are discovered, one typically does a `gridding'
observation to refine the position (as the initial survey position can
be quite uncertain, e.g.\ $\pm7$~arcmin for the PMPS). This facilitates
observations at higher radio frequencies and also obtaining a timing
solution. However, the interpretation of gridding observations is
difficult for RRATs (i.e. in the case of a non-detection it is unclear
whether one was off source or whether the RRAT was `off'). Therefore,
refining positions must typically wait for a phase-connected timing
solution, requiring a year's observation to remove the effects of the
Earth's motion around the Sun.  While most pulsars are timed through
their average profiles, RRATs are timed from their isolated pulses and
often a large amount of time may be required to accumulate enough
pulses to achieve a solution.  Consequently, some RRATs (especially
those with the lowest detected burst rates) still do not have
solutions after several years and some are so sporadic that it is not
possible to determine a timing solution.
When solutions are determined, errors on timing parameters can be
much higher than for normal pulsars due to pulse-to-pulse jitter and varying numbers of components in the single pulses
 \citep{lmk+09,hey+11}.

\begin{figure}
  \begin{center}
    \includegraphics[
      angle=-90,width=\textwidth]{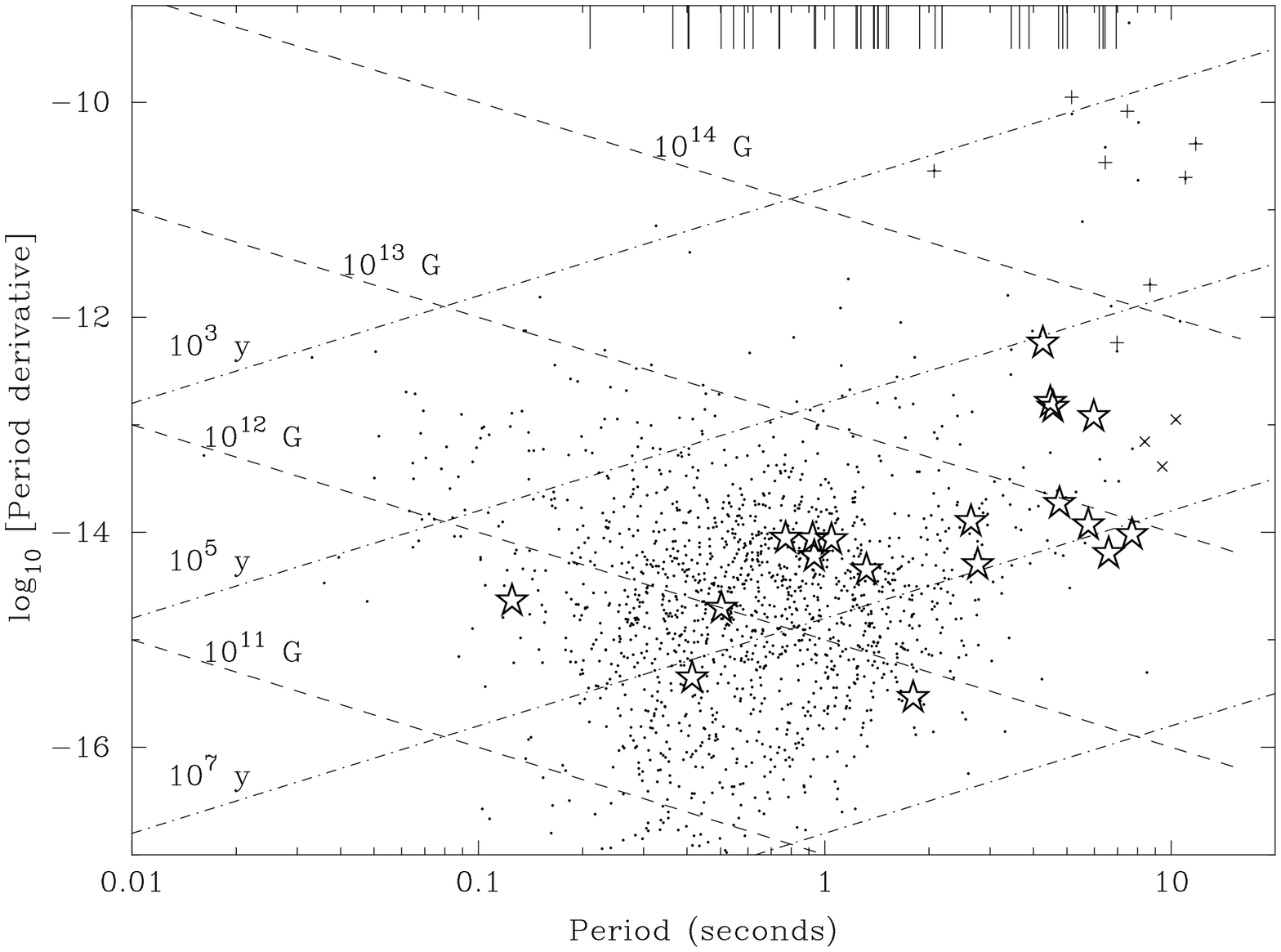}
  \end{center}
  \caption{\small{A portion of the pulsar $P-\dot{P}$ diagram showing
      all known pulsars. The 19 sources (18 PMPS `RRATs' \citealt{mlk+09,kkl+11} and one GBT 350-MHz RRAT \citealt{bmm11})
      which have determined period derivatives are shown as open black stars.
      Magnetars are marked as plus signs and XINS as crosses.  Also
      over-plotted are lines of constant `characteristic age' and
      vacuum dipole estimate of the magnetic field
      (see
      e.g.\ \citet{lk05} for definitions of all of these quantities). The periods of the RRATs with no measured period derivatives are
shown at the top of the diagram.}}
  \label{fig:ppdot}
\end{figure}

Through observations with the Parkes, Lovell, Green Bank, and Arecibo
telescopes, timing solutions have been determined now for 19 PMPS
RRATs~\citep{mll+06,mlk+09,lmk+09,kkl+11,bmm11}.
Achieving phase-connected solutions allows us
to place
these sources on the period-period derivative diagram, a commonly used
scatter plot for pulsar classifications, as in
Fig.~\ref{fig:ppdot}. These timing solutions have shown that there
is no evidence of any binary companions for these sources.
As can be seen from
Fig.~\ref{fig:ppdot}, while the spin-down properties of RRATs and other pulsars overlap,
RRATs in general have longer periods and higher magnetic
fields than other pulsars. A Kolmogorov--Smirnov test shows probabilities of $1\times10^{-12}$
and $2\times10^{-7}$ that the periods and magnetic fields of
RRATs are drawn from the same population as normal non-recycled pulsars, respectively. These probabilities are lower
than calculated in \citet{mlk+09} from a sample of nine RRATs, indicating that this trend is robust.
However, it is unclear how much of this is intrinsic and how much is due to the selection effects of surveys.
One interesting observation is that, while many RRATs populate the same region of $P-\dot{P}$ space as
other pulsars, several RRATs occupy $P-\dot{P}$ space devoid of normal pulsars and closer to the population of XINS. This could imply an
evolutionary relationship between these objects, as suggested by \citet{ptr06}.

Timing solutions have also
facilitated high-energy follow-up observations (discussed below), and
enabled monitoring for timing irregularities. In particular, they
allowed the detection of two glitches from
PSR~J1819$-$1458~\citep{lmk+09}. The glitch recovery is anomalous in
that it results in a net \textit{decrease} in the star's slow-down
rate, the opposite trend to that in all other pulsar
glitches~\citep{elsk11}. \citet{lmk+09} speculate that if this
behaviour were typical for PSR~J1819$-$1458 it may have evolved from
the magnetar region of the $P-\dot{P}$ diagram. This source is being
monitored regularly with the Lovell telescope in the UK and with the
Urumqi telescope in China~\citep{ezy+08,hey+11}. No further glitches
have yet been seen. No glitches have been detected as yet from any of
the other sources. PSR~J1819$-$1458 is also the only source which has
measured radio polarisation properties. \citet{khv+09} found it to be
similar to other slow, low spindown-energy pulsars and is seen to have
a `normal' S-shaped swing of the polarisation position angle.

A recent analysis by \citet{pmk+11} of eight of the original 11 PMPS
RRATs searched for short (approximately hours) and long-term (weeks to years)
periodicities in the detected pulse arrival times.   They found significant periodicities in the
pulse arrival times for six RRATs with timescales ranging from minutes to years and were not found in sequences of simulated
random pulses. They also found that for
the two brightest and most prolific sources (J1317$-$5759 and
PSR~J1819$-$1458) there were more instances of consecutive pulses seen
than would be expected from a random distribution of pulse times (though most pulses, even for these sources,
are solitary). Another analysis of RRAT spectra based on multi-frequency radio observations by \citet{mmr+11}
showed that RRATs have typical
steep pulsar spectra with spectral indices $\sim$1.7~\citep{mmr+11}
unlike the flat spectra seen in radio-emitting
magnetars~\citep{crh+06,crj+08}. We note, however, that these spectral indices are based on measurements at frequencies
$>$700~MHz and that lower-frequency spectral turnovers are possible.

\subsection{Optical and Infrared}
PSR~J1819$-$1458 has been observed in the optical in the search of a
counterpart to the radio and X-ray (see below) source by
\citet{dml06}. They observed this source with ULTRACAM~\citep{dms+07}
on the 4.2-metre William Herschel Telescope, acquiring $97100$ images,
each with an $\sim$18~ms exposure time. No detection of any optical
bursts were made with a limiting burst magnitude of $i'=16.6$. To
improve upon this, a follow-up study again using the William Herschel
Telescope as well as on the New Technology Telescope was performed in
tandem with simultaneous radio observations with the Lovell Telescope
at Jodrell Bank~\citep{dkm+11}. The goal was to identify the times
when radio pulses were detected and examine the corresponding optical
frames. This is the best way to obtain an optical limit under the
assumption that the optical emission is also transient and connected
to the radio emission, as seen in the Crab~\citep{sso+03}. No evidence
for optical bursts or increased optical flux during radio bursts was
found. Besides these observations of PSR~J1819$-$1458 there have been
no other reported optical observations of RRATs.

\citet{rct+10} attempted to detect infrared counterparts of the first
three RRATs with timing solutions, and hence the first with accurately
determined positions. Using the Very Large Telescope in Chile they
found no infrared counterpart to J1317$-$5759 down to a limiting
magnitude of $K_{\mathrm{s}}\sim 21$. For J1913$+$1330 the
observations used an incorrect source position due to a slight (yet
crucial) inaccuracy in the originally published timing solution, so
that the source lay outside of their field of view. A putative
counterpart is reported for PSR~J1819$-$1458 on the edge of the 1-$\sigma$
error circle of the X-ray position, but \citet{rct+10} conclude that
the probability of a chance alignment is high.

\subsection{X-ray}
Several X-ray observations of PSR~J1819$-$1458 have been made, at
first a serendipitous detection by \citet{rbg+06} using
\textit{Chandra} (30~ks) and then a targeted observation with
\textit{XMM Newton} (43~ks) by \citet{mrg+07a}. These revealed a
thermal spectrum with $kT$$\sim$140~eV, characteristic of a cooling
neutron star. Pulsations at the $4.2$-s radio-determined period were
also observed, as well as a $0.5$~keV spectral absorption feature,
tentatively identified with proton cyclotron resonant scattering. Such
features enable an independent estimate of the magnetic field. The
inferred magnetic field strength from the cyclotron interpretation is
$2\times10^{14}$~G. Equating this with the (commonly used but
incorrect) vacuum dipole estimate for the magnetic field predicts the
magnetic and rotational axes to be misaligned by
$\alpha=15\degree$. Interestingly, equating the cyclotron estimate
with that from the force-free model of \citet{spi06} yields
$\sin^2\alpha < 0$, meaning either the cyclotron interpretation or the
force-free magnetic field strength estimate is incorrect. The spectral
feature is confirmed in {\it Chandra} observations (30~ks) made by
\citet{rmg+09} which also reveal evidence for extended emission up to
$5.5$~arcsec from the star. \citet{rmg+09} suggest that this emission
is due to a pulsar wind nebula but cannot explain the power source
given what would be a large required X-ray efficiency (see e.g.\
\citealt{gs06}). They suggest that the PWN could be powered by the large
magnetic field of this RRAT. A detailed study of PSR~J1819$-$1458
using \textit{XMM Newton} ($94$~ks) in tandem with three radio
telescopes has reported initial results confirming the earlier
findings~\citep{mmr+11}, with a comprehensive analysis, including a
correlation analysis of radio/X-ray flux, in preparation.

In addition to PSR~J1819$-$1458, PSR~J1317$-$5759 was observed with
\textit{XMM-Newton} ($32$~ks) but with no
detection~\citep{rm08,mac09}. PSR~J1913$+$1330 was also observed,
using \textit{Swift} ($9.3$~ks), but again with an inaccurate source
position~\citep{rm08,mac09,rct+10}. When improved astrometry became
available for PSR~J0847$-$4316 and PSR~J1846$-$0257~\citep{mlk+09},
these were both observed with \textit{Chandra} ($10.7$~ks and
$196.1$~ks respectively) but neither were detected, with the
corresponding limits obtained being reported in \citet{kec+09}. An
X-ray candidate for PSR~J1911$+$00 has also been suggested by
\citet{hrf+06} which fortuitously occupies the same field as Aql X-1,
a much-observed low-mass X-ray binary, but no pulsations were detected
and the spectrum is not thermal. With by far the lowest rate of
detected bursts of the original PMPS RRATs, PSR~J1911+00 is unlikely
to ever have an accurate position determined, and hence progress in
identifying this X-ray source is unlikely.

\subsection{Gamma-ray}

With the advent of the {\it Fermi} satellite, we now
know of nearly 100 rotation-powered pulsars that are detectable at gamma-ray energies.
The gamma-ray luminosity is roughly proportional to the spin-down luminosity, with all
of the {\it Fermi} sources in the first catalog having $\dot{E}$s greater than $10^{34}$~erg~s$^{-1}$. Only one RRAT, J1554$-$5209, has $\dot{E}$ greater
than  $10^{34}$~erg~s$^{-1}$; \citet{kea10a} folded {\it Fermi} photons using the radio ephemeris and found no evidence for gamma-ray emission from
this source.
The remaining 18 RRATs have $\dot{E}$s less than $10^{33}$~erg~s$^{-1}$ and are not predicted to be gamma-ray sources, though \citet{kea10a} detect
 a folded profile of 4-$\sigma$ significance with {\it Fermi} for J1819$-$1458. However,
they only detect emission in half of the {\it Fermi} dataset.
A longer timespan of {\it Fermi} observations is necessary to gauge the reality of this signal.

\section{Where do these sources fit in?}\label{sec:discussion}
With the discovery of so many transient radio neutron stars in single
pulse searches\footnote{For a complete list of the pulsars which have
  been found in single pulse searches since 2006 refer to the website
  \texttt{http://www.as.wvu.edu/\string~pulsar/rratalog/}.} we now know
of RRATs covering a large range in $g$, the fraction of the observed
pulse periods where we do not detect a pulse (so that $1-g$ is an
upper limit on the nulling fraction). The pulsars identified as RRATs
span the range $10^{-4}\lesssim g \lesssim 10^{-1}$ and, as noted by
\citet{bbj+11}, if we assume these sources are indeed nulling, then
they seem to be an extension of the nulling pulsars discovered in the
PMPS by \citet{wmj07}.

As well as in nulling fraction, the distribution seems to be
continuous in the typical `off' timescales for these `nullers' and
`RRATs', spanning the range $10\,\mathrm{s} \lesssim t_{\mathrm{off}}
\lesssim 10^4\,\mathrm{s}$~\citep{bbj+11}. This hints at a continuum
in intermittency timescales, but beyond hour-long timescales there is
a dearth of known objects until we reach the so-called `intermittent
pulsars', PSRs~B1931$+$24, J1832$+$0029, and J1107$-$5907 where the
relevant timescales range from $\sim$1~week to
$~1$~year~\citep{klo+06,kra08}. Of course the lack of objects with
such long `off' timescales is to be expected due to the difficulty in
discovering (and confirming\footnote{In general a pulsar is not
  considered to be a truly new discovery until it has been confirmed
  in at least one follow-up observation, post-survey.}) such sources.

\subsection{The trouble with timescales}
The timescales on which pulsars exhibit variability span more than 16
orders of magnitude, from the nano-structure seen in the Crab pulsar
($\sim$$10^{-9}$~s) to the approximately year-long timescales of intermittent
pulsars ($\sim$$10^7$~s). However it is very difficult to explain such a
large range of timescales.
It is an open question as to how stable pulsar emission can occur (at
all!) and for days or weeks before apparently switching off. But the
picture is even more complicated as \textit{periodic} switching
\textit{back and forth} between (at least two) stable magnetospheric
states has been reported in six pulsars studied by \citet{lhk+10}. They
report correlated period derivative and pulse shape changes which,
when understood to be the result of a switch between two stable
states, can remove the so-called `timing noise' in these pulsars. Such
results hint that observed phenomena like nulling (or the more general
phenomena of mode-changing, and therefore RRATs), pulse-shape changes
and timing noise are all one and the same and a result of
magnetospheres switching between stable states.

A number of stable solutions have been obtained numerically for
force-free magnetospheres \citep{ckf99,tim06}. Switching between these
stable configurations is possible if there is a sudden depletion of
charges in the magnetosphere. \citet{con05}, \citet{tim09}, and
\citet{rmt11} have even discussed how these switches would alter the
radio emission and result in observed moding/nulling, but no trigger
mechanism for such an event is known. Perhaps the most puzzling aspect
of all is that the mechanism for these switches is quasi-periodic in
nature. One possible cause of quasi-periodicities could be due to
modulations of magnetospheric particle density by asteroids orbiting
the neutron star \citep{cs08}.  However, due to the low mass needed in
these belts, they may be undetectable through reflected radio emission
until the SKA or directly without a larger infrared telescope.

\subsection{The population of radio transients}
We note that the large projected population of pulsars likely to be
detected as RRATs may not be a cause for concern. This is because it
seems that many can be accounted for as weak/distant sources with high
modulation indices. If the undetected underlying emission from such
pulsars is above the typical minimum radio luminosity above which the
pulsar population is estimated ($L_{\mathrm{min}}\sim
0.1\,\mathrm{mJy}\, \mathrm{kpc}^2$) then these pulsars are already
accounted for within the scaling factors used to calculate the
estimated total number of pulsars (above $L_{\mathrm{min}}$) from the
observed sample (see e.g.\ \citealt{lfl+06,rl10a}). However, if some of
the RRATs are indeed nulling pulsars, or pulsars whose underlying
emission is below the arbitrarily-imposed luminosity cut-off (see
e.g.\ \citealt{fk06}), then they will indeed contribute to the number of
Galactic pulsars. Pulsar population estimates do not take account of
nulling when determining the likelihood of detection of pulsars of
various types (which consequently inform how known selection effects
are removed to reach an estimate for the total population), another
contribution (in addition to a luminosity-limited calculation) to the
fact that the total number of active pulsars in the Galaxy should be
larger than estimated through these methods.

The search for radio transients has also resulted in the discovery of
other interesting sources. Most notably, as we have mentioned above,
is the Lorimer Burst, of which there may now be
two~\citep{lbm+07,kkl+11}. As radio frequency interference is a major
hindrance to such searches, it is not surprising that previously
unknown types of terrestrial signals have also been identified in
transient searches. \citet{bbe+11} report the detection of signals,
referred to as `perytons', with anomalous dispersion characteristics
which are not of astrophysical origin\footnote{They are detected in
  all 13 beams of the Parkes 21-cm multi-beam receiver, which, as
  \citet{bbe+11} note, is impossible for a boresight signal.}. Such
terrestrial signals are a good illustration of the limits of
single-dish radio telescope surveys. This is another argument in
favour of the move towards radio arrays where signals from many
elements (not necessarily dishes) can be correlated as appropriate to
identify the source of terrestrial signals and to remove them.

\subsection{RRATs with TNG instruments}
Many of the next generation of radio telescopes which will be used for
pulsar and fast transient searches are array instruments. The
potential of transient searches has now truly been recognised, with
all of these major telescopes (e.g.\ LOFAR, FAST, ASKAP, MeerKAT, SKA)
having pulsars and fast transients as key science
objectives~\citep{jtb+08,bbjf09,leu11,nlj+11}. LOFAR, for instance, currently in its
commissioning stage, but is already providing scientific results in
the field of pulsars~\citep{sha+11}. LOFAR predicts the discovery of
$\sim$1000 pulsars in an all-sky survey \citep{ls10}; possibly hundreds of RRATs will be discovered.
 Even greater
advances will come with surveys for transients in interferometric
images on short time scales.
 They will enable accurate positional localisation at the
time of discovery, facilitating timing solutions and immediate
high-energy follow-up. As a result of the transition to `software
telescopes' like LOFAR, much software development is needed. Already,
much work is underway in, e.g.\ live transient searches using
graphical processor units as recently demonstrated by
\citet{mks+11}. Substantial advancements like this are essential to
enable us to deal with data from the SKA by the end of the decade.

\section*{Acknowledgements}
EK acknowledges the FSM, as well as Joris Verbiest and Michael Kramer
for useful comments. MAM is supported by the NSF, WVEPSCOR, and
by the Research Corporation.


\label{lastpage}
\end{document}